\documentclass[journal]{IEEEtran}
\IEEEoverridecommandlockouts
\usepackage{cite}
\usepackage{amsmath,amssymb,amsfonts}
\usepackage{algorithmic}
\usepackage{hyperref}
\usepackage{graphicx}
\graphicspath{{figures/}}
\usepackage{textcomp}
\usepackage{xcolor}
\usepackage{color}
\usepackage{booktabs}
\usepackage{enumitem}

\def\BibTeX{{\rm B\kern-.05em{\sc i\kern-.025em b}\kern-.08em
    T\kern-.1667em\lower.7ex\hbox{E}\kern-.125emX}}

\begin{document}

\title{Non-destructive Characterization of Anti-Reflective Coatings on PV Modules
\thanks{The project was primarily funded and intellectually led as part of the Durable Modules Consortium (DuraMAT), an Energy Materials Network Consortium funded under Agreement 32509 by the U.S. Department of Energy (DOE), Office of Energy Efficiency \& Renewable Energy, Solar Energy Technologies Office (EERE, SETO). Lawrence Berkeley National Laboratory is funded by the DOE under award DE-AC02-05CH11231. This work was authored in part by the NREL, operated by Alliance for Sustainable Energy, LLC for the US DOE under contract no. DE‐AC36‐08GO28308.
\newline
\newline
The authors declare no conflicts of interest. The views expressed in the article do not necessarily represent the views of the DOE or the U.S. government. Instruments and materials are identified in this paper to describe the experiments. In no case does such identification imply recommendation or endorsement by LBL or NREL. The U.S. government retains and the publisher, by accepting the article for publication, acknowledges that the U.S. government retains a nonexclusive, paid-up, irrevocable, worldwide license to publish or reproduce the published form of this work, or allow others to do so, for U.S. government purposes.
}
}

\author{
\IEEEauthorblockN{Todd Karin~\IEEEauthorrefmark{1}, David Miller~\IEEEauthorrefmark{2}, Anubhav Jain~\IEEEauthorrefmark{1} \vspace{15pt}} \\
\IEEEauthorblockA{
\IEEEauthorrefmark{1}Lawrence Berkeley National Laboratory, Berkeley, CA, U.S.A 
}\\
\IEEEauthorrefmark{2}National Renewable Energy Laboratory, Golden, CO, U.S.A 
}

%
%\author{\IEEEauthorblockN{Todd Karin}
%\IEEEauthorblockA{\textit{Lawrence Berkeley National Laboratory} \\
%Berkeley, CA, USA \\
%toddkarin@lbl.gov}
%\and
%\IEEEauthorblockN{Anubhav Jain}
%\IEEEauthorblockA{\textit{Lawrence Berkeley National Laboratory} \\
%Berkeley, CA, USA \\
%ajain@lbl.gov}
%}

\maketitle

\begin{abstract}
Anti-reflective coatings (ARCs) are used on the vast majority of solar photovoltaic (PV) modules to increase power production. However, ARC longevity can vary from less than 1 year to over 15 years depending on coating quality and deployment conditions. A technique that can quantify ARC degradation non-destructively on commercial modules would be useful both for in-field diagnostics and accelerated aging tests. In this paper, we demonstrate that accurate measurements of ARC spectral reflectance can be performed using a modified commercially-available integrating-sphere probe. The measurement is fast, accurate, non-destructive and can be performed outdoors in full-sun conditions. We develop an interferometric model that estimates coating porosity, thickness and fractional area coverage from the measured reflectance spectrum for a uniform single-layer coating. We demonstrate the measurement outdoors on an active PV installation, identify the presence of an ARC and estimate the properties of the coating.
\end{abstract}
%
%\section{Outline}
%-  Motivation/intro. 
%
%- Interferometry to measure coating thickness. Comparison of assembled module vs. glass coupon ARC.
%
%- This measurement can be performed on assembled modules. See reflection from cell, but from 450-1000 nm, have good measurement.
%
%- Figure comparing reflection spectrum on glass coupon with oil + MC silicon cell below, coupled with microscope oil, to backside spray painted black. This demonstrates that the integrating sphere method is accurate in the region of interest. Also MOB sample.
%
%- Theory can be used to determine coating thickness and porosity (assuming that silica is used). ARC model. Plots showing Reflection vs. wavelength for a variety of coatings.
%
%- Probe design. A smaller integration area lowers reflection from cell. The 45-45 measurement is even better at getting rid of the cell, but has the problem that a diffusely reflecting sample would appear to be a better coating. 
%
%- Comparison of reflection and transmission on David Miller samples. Use color theory to predict coating porosity and thickness from reflection images. 
%
%- Measurement on cored samples?
%
%- Reflection imaging is also useful. Color shift is related to thickness. Simple visual with a cellphone camera (Separate paper?)
%
%- Compare to handheld spectrophotometer. 
%
%- Keep the paper not too long.

\section{Introduction}

\IEEEPARstart{A}{nti}-reflective coatings (ARCs) on the air-glass interface of solar photovoltaic (PV) modules are a cost-effective way to improve power production. In 2019, ARCs were used on over 90\% of commercial PV modules, typically resulting in approximately 3.0\% more light delivered to the solar cell \cite{ITRPV2020,DSM3percent}. Many of these coatings consist of a $\sim$125~nm layer of porous silica deposited by a sol-gel process~\cite{Glaubitt2012, Freiburger2017}. 

In addition to graded-index coatings (e.g., constructed of porous silica) currently in widespread deployment, recent work has studied the ability of multi-layer dielectric coatings  to improve transmission~\cite{Law2020} and thermal management~\cite{Slauch2018} of PV modules.  The use of fully-dense metal-oxide materials may give greater robustness relative to graded-index coatings. For example, multi-layer coatings composed of hard materials with a high mechanical modulus have been found to be significantly more durable to abrasion~\cite{Newkirk2021}. Because there are no pores, the permeability of water is reduced, preventing moisture ingress and subsequent damage in the event of glass corrosion~\cite{Clark1979} of the underlying substrate and/or additional strain in the event of freezing. 
%The insulating characteristics typical to metal oxides can also limit the transfer of charge, as in the event of potential induced degradation~\cite{Hacke2018}.

Although the single-layer ARCs on the market today can have an expected lifetime of 15 years~\cite{ITRPV2020}, ARC lifetime depends on the coating deposition process \cite{Simurka2018} and deployment conditions including heat, humidity, soiling and frequency/method of cleaning. For example, total ARC loss has been observed after just 1 year in desert environments~\cite{Miller2020}. Depending on the extent of other degradation modes, such a limited ARC lifetime push a solar module out of warranty.

Low ARC lifetimes also impact the accuracy of energy production predictions. While the PV community often assumes a linear module power degradation rate of 0.6\% per year~\cite{Jordan2016}, if the power enhancement of 3\% due to the ARC is lost over several years then there can be large errors in long-term energy-yield predictions. The power loss could be even greater than 3\% since porous silica coatings can also provide anti-soiling benefits~\cite{Einhorn2019,Glaubitt2012}. Thus, understanding coating degradation is crucial to predicting long-term performance. 

Single-layer ARCs based on porous silica often have tradeoffs between performance and durability~\cite{Ilse2019,Ganjoo2019}. While coating performance improves as the porosity increases, higher porosity can result in lower coating hardness and wear resistance~\cite{Simurka2018, Pop2014}. Although coating manufacturers use a number of stress tests and technical solutions to develop better coatings~\cite{Buskens2009, Ganjoo2019}, third-party module reliability testing programs do not measure ARC longevity \cite{PVEL_PQP}. This unfortunately creates a short-term financial incentive for a module manufacturer to use a less-durable but higher-performance coating.

Many prior studies have used optical techniques to quantify coating performance~\cite{Moffitt2020, Yoldas1980, Nielsen2015, Galy2020}. Other work tested ARC longevity under artificial or natural aging~\cite{Ilse2019, Miller2020, Pop2014}.  Most of this work uses glass coupons and optical transmittance measurements, a straightforward and reliable method, but one that is not applicable to assembled modules. While reflectance measurements have been performed on assembled modules~\cite{Parretta1999, Maddalena2003}, the methods require complex custom equipment. New techniques are needed to test coating reliability under accelerated aging and to quantify coating performance on fielded PV modules.

In this paper, we demonstrate how to measure ARC reflectance on commercial PV modules, expanding on recent work by the authors~\cite{Karin2020pvsc}. For PV glass measurements, current standards recommend a large integrating sphere (diameter $>$150~mm) which typically has a large spot size (diameter $>$8~mm) to collect all angles of reflected light and achieve spatial averaging~\cite{IEC-62805-1,IEC-62805-2}. However, we find that to accurately measure ARC reflectance on an assembled module, it is important to minimize the interrogation area to 1~mm diameter in order to remove reflection from the PV cell. This paper makes clear the need for a new standard for measuring air-glass reflectance from assembled PV modules.

The methods in this paper are fast ($<$1 second per acquisition, $<$ 1 hour for set up on site), do not require any electrical reconfiguration of the PV module and use commercially-available equipment with simple modifications. A variety of commercial optical probes can be used, but probes with higher light intensity allow the measurement to be performed accurately outdoors in full-sun conditions.

For single-layer ARCs, we model the reflectance using the transfer-matrix form of the Fresnel equations~\cite{TMM} and use numerical optimization to predict the thickness, porosity and area fraction remaining of the coating~\cite{Galy2020}. We describe two suggested figures of merit: solar-weighted photon reflectance and nominal power enhancement. The calculations performed in this paper are available in an open-source python package~\cite{pvarc}.

\section{Methods for Measuring ARCs on PV Modules}\label{sec:methods}

\begin{figure}
\centering
\includegraphics{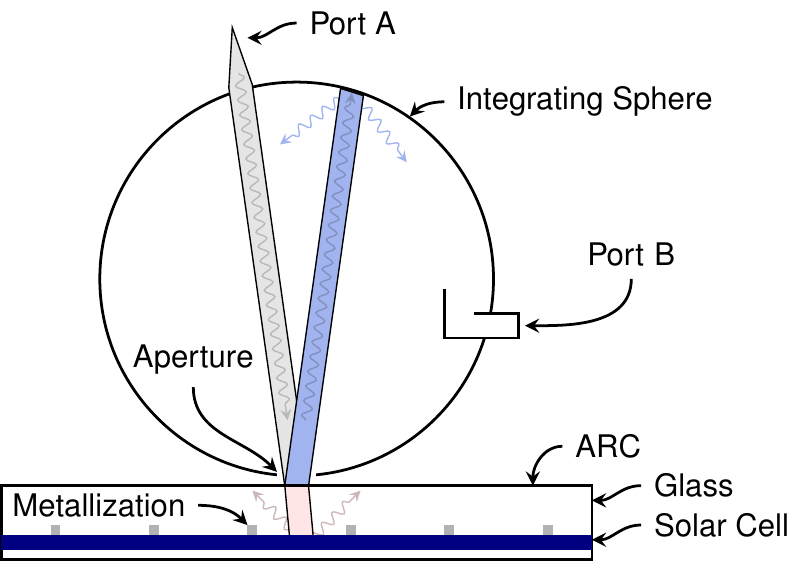}
\caption{Integrating sphere probe for measuring ARC spectral reflectance. The aperture size controls how much light reflected from the cell returns into the integrating sphere. Two possibilities for measuring reflectance are possible. In the ``excite-8$^\circ$, collect all angles" geometry, broadband light is injected into port A and focused to a small spot size on the sample. Reflected light fills the integrating sphere and is sampled from measurement port B. In the ``excite all angles, collect 8$^\circ$" geometry, light is instead injected into port B and sampled from port A. The Helmholtz reciprocity theorem shows that these two optical geometries produce the same result.}
\label{fig:diagram}
\end{figure}

Current standards for measuring reflectance and transmittance for solar module glass, IEC 62805-1 and 62805-2, are not designed to accurately measure air-glass reflectance from an assembled module~\cite{IEC-62805-1, IEC-62805-2}. If applied to an assembled module, light reflected from the PV cell, metallization and/or backsheet is also collected in the measurement (Fig.~\ref{fig:diagram}), leading to a systematic overestimate of reflectance from the ARC.

\begin{figure}
\begin{center}
\includegraphics{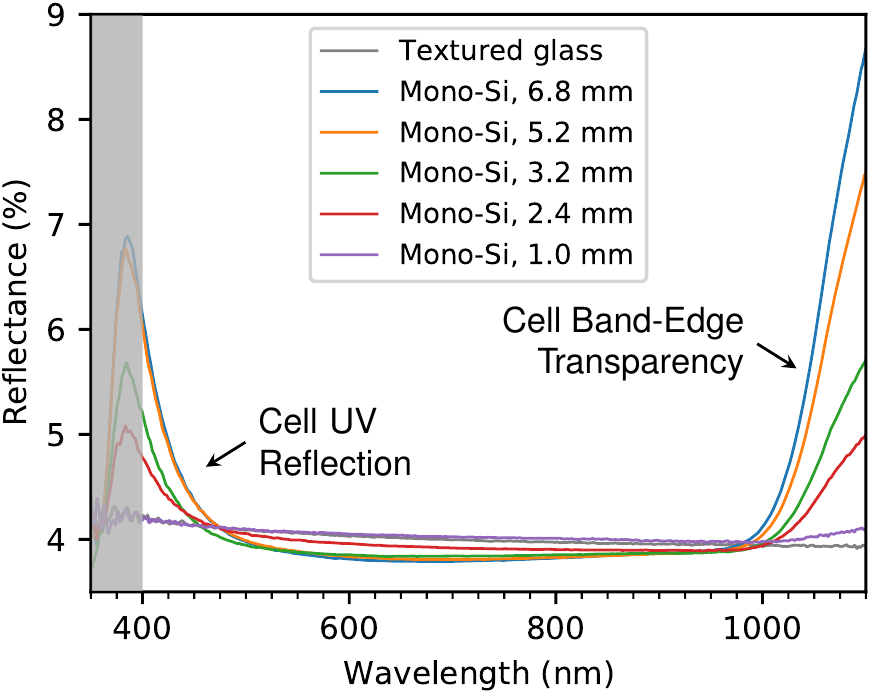}
\caption{A smaller diameter aperture on the integrating sphere results in a reduced impact of the cell on the measured reflectance spectrum. Measurements are taken with the probe centered between two grid lines on a mono-Si minimodule without an ARC. The cell has a grid spacing of 2.5 mm. By reducing the aperture to 1.0~mm inner diameter, the valid measurement range is extended to 400 through 1000 nm. The textured glass sample has the back blackened and uses a 1.0~mm aperture. Below 400~nm, the tungsten halogen light source does not provide enough light intensity to accurately measure reflectance due to stray light in the spectrometer.}
\label{fig:effectofaperture}
\end{center}
\end{figure}

Since silicon solar cells are highly textured to promote absorption, reflection from the cell is diffuse. Thus, reducing the aperture size should minimize light collected from interfaces below the glass surface. We test this idea with reflectance measurements on a mono-Si mini-module using different aperture sizes, centering the probe over a spot between two grid-lines (further details in Appendix~\ref{app:aperture}). The resulting reflectance spectra in Fig.~\ref{fig:effectofaperture} show two contributions: a $\sim$4\% reflectance from the air-glass interface, a UV reflection at 400 nm from the top surface of the Si cell, and a band-edge reflection from the PV cell at 1100~nm. The band-edge reflection occurs because the cell becomes transparent below the band gap and light is reflected from the back of the cell. The shape of the curve is well-explained by the external quantum efficiency (EQE) of a standard Si cell~\cite{Green2020}. 

Figure~\ref{fig:effectofaperture} shows that smaller apertures strongly reduce the band-edge and UV contribution to the total reflectance spectrum. The impact of a reduced aperture can be estimated by inspecting the optical configuration in Fig.~\ref{fig:diagram}. The fraction of returned light into the integrating sphere is

%\begin{equation}\label{eq:dep}
%\begin{split}
%R_\text{total}(\lambda) = R_\text{air-glass}(\lambda) + & f_\text{cell} [T_\text{glass}(\lambda) T_\text{enc}(\lambda)]^2  R_\text{enc-cell}(\lambda) \\  & +  f_\text{enc} [T_\text{glass}(\lambda)]^2  R_\text{glass-enc}(\lambda)
%\end{split}
%\end{equation} 
\begin{equation}\label{eq:dep}
R_\text{total} = R_\text{air-glass} +  f_\text{cell} T_\text{glass}^2 T_\text{enc}^2  R_\text{enc-cell}  +  f_\text{enc} T_\text{glass}^2  R_\text{glass-enc}
\end{equation} 
where  \(R_\text{air-glass}\) is the hemispherical reflectance from the air-ARC-glass stack, \(R_\text{enc-cell}\) is the reflectance from the encapsulant-cell interface, \(R_\text{glass-enc}\) is the reflectance from the glass-encapsulant interface, \(f_\text{cell}\) and \(f_\text{enc}\) are wavelength-independent geometrical factors and \(\lambda\) is the wavelength.  The third term in Eq.~\ref{eq:dep} due to reflection from the glass-encapsulant interface is typically small and therefore we ignore it in the following. The transmittance values are given by
%\begin{align}\label{eq:Rtot}
%T_\text{glass}(\lambda) & = [1-R_\text{air-glass}(\lambda)] [1- \alpha_\text{glass}(\lambda)] \\
%    T_\text{enc}(\lambda) & = [1-R_\text{glass-enc}(\lambda)] [1- \alpha_\text{enc}(\lambda)] 
%    \end{align}
    \begin{align}\label{eq:Rtot}
T_\text{glass} & = (1-R_\text{air-glass})(1- \alpha_\text{glass}) \\
    T_\text{enc} & = (1-R_\text{glass-enc})(1- \alpha_\text{enc}) 
    \end{align}
where \(\alpha_\text{glass}\) is the absorption in the glass and $\alpha_\text{enc}$ is the absorption in the encapsulant.

The geometrical factor $f_\text{cell}$ can be estimated by assuming reflection from the encapsulant-cell interface is isotropically diffuse, a good approximation for modern highly-textured solar cells. The amount of light returning into the integrating sphere is then approximately equal to the solid angle of the aperture viewed from the cell divided by the solid angle of a hemisphere:
\begin{equation}\label{eq:f}
f_\text{cell} \approx \frac{2}{1 + \left[2 (h_\text{glass} + h_\text{enc})/d\right]^2 },
\end{equation}
where \(d\) is the diameter of the aperture, \(h_\text{glass}\) is the thickness of the glass and \(h_\text{enc}\) is the thickness of the encapsulant. Equation~\ref{eq:f} approximates the input light as a single ray at the center of the aperture at near-normal angle-of-incidence, the air-glass interface as flat, and the air-glass interface transmission as specular.

We can test this theory using the spectral reflectance measured in Fig.~\ref{fig:effectofaperture}. The measured value of the second term on the right hand side of Eq.~\ref{eq:dep} is found by subtracting the textured glass reflectance spectrum from the total reflectance spectrum in Fig.~\ref{fig:effectofaperture} and averaging from 1020 to 1100~nm, the region where $R_\text{enc-cell}$ is high enough to be measured. The ratio of this second term for spectra with different apertures should be proportional to \(f_\text{cell}\) given by Eq.~\ref{eq:f}. We compare the theory and experiment by scaling the experimental values to the value of \(f_\text{cell}\) at $d=6.8~\text{mm}$ and using a thickness of $3.175~\text{mm}$ for the glass and 0.45~mm for the encapsulant. The excellent agreement with Eq.~\ref{eq:f} and the experimental values in Table.~\ref{tab:aperture} confirms how small apertures reduce the contribution of cell reflectance to the total reflectance spectrum.

\begin{table}
\caption{Comparison of theoretical and measured geometrical factor \(f_\text{cell}\) describing the contribution of reflection from the cell to the total reflectance spectrum.}
\label{tab:aperture}
\begin{center}
\begin{tabular}{lrr}
\toprule
d  (mm) &  $f_\text{cell,theory}(d)$ &   $f_\text{cell,exp}(d) \cdot\frac{f_\text{cell,theory}(d=6.8\,\text{mm})}{f_\text{cell,exp} (d=6.8\,\text{mm})} $ \\
\midrule
              6.80 &      0.94 &                            0.94 \\
              5.20 &      0.68 &                            0.69 \\
              3.20 &      0.33 &                            0.34 \\
              2.42 &      0.20 &                            0.20 \\
              1.03 &      0.04 &                            0.04 \\
\bottomrule
\end{tabular}
\end{center}
\end{table}

Therefore, we conclude that one way to obtain an accurate measurement of ARC reflectance is to reduce the aperture diameter so that the second and third terms in Eq.~\ref{eq:dep} can be ignored.

We can make an estimate of the diameter needed to achieve a 0.1\% absolute accuracy. The reflectance from a typical mono-Si cell is lower than 2\% within the range of 475 to 1000 nm~\cite{Gangopadhyay2013} (accounting for difference between a measurement in air vs. under glass). If ${R_\text{enc-cell} \leq 2.0\%}$ and diffuse, glass thickness is 3.2~mm and encapsulant thickness is 0.5~mm, then an aperture \({d\leq1.2~\text{mm}}\) makes the second term on the right-hand-side of Eq.~\ref{eq:dep} smaller than the desired accuracy of 0.1\% within the range of 475 to 1000 nm.

\begin{figure}
\centering
\includegraphics{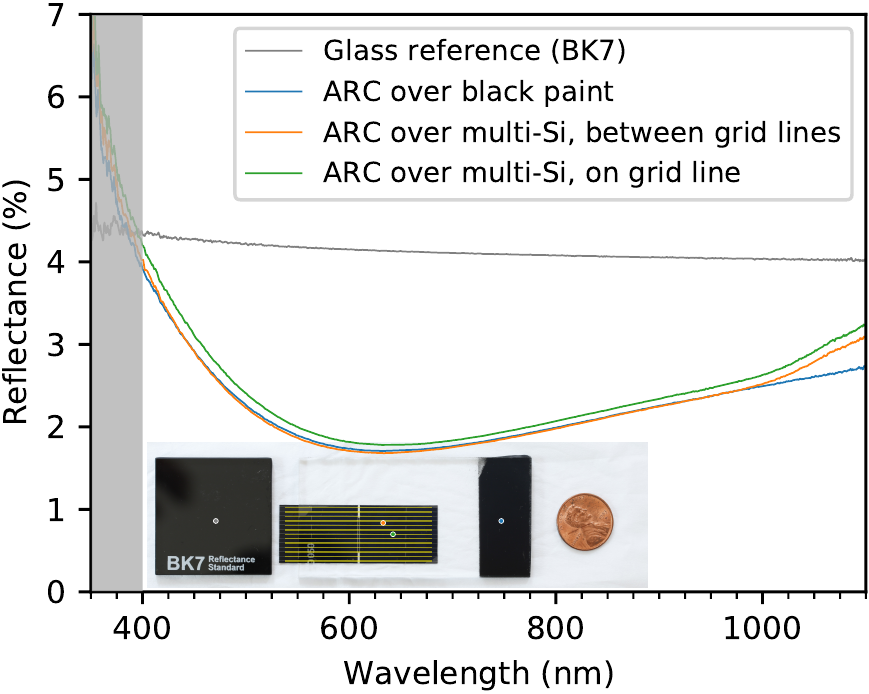}
\caption{Reflectance measured over black paint or a multi-Si PV cell is the same within 400-1000 nm when the probe is centered between grid liens. In contrast, when the probe is centered on the grid lines, a slightly increased reflectance is observed. Inset shows samples and measurement locations including (from left to right) BK7 glass reference (back painted black), ARC-coated glass slide with multi-Si cell beneath (coupled with microscope oil) and a section of the ARC with back painted black. Grid lines on cell are highlighted in yellow. Measurement taken with a 1.0~mm aperture in the excite 8$^\circ$, collect all angles geometry.}
\label{fig:effectofcell}
\end{figure}

For PV technologies with sun-side metallization, it is important to center the integrating sphere probe between grid lines~\cite{Parretta1999}. To demonstrate the importance of measurement location, we place a small multi-Si cell underneath a glass coupon with an ARC (unaged sample D from \cite{Miller2020}), coupled with \({n=1.5}\) microscope oil, see inset of Fig.~\ref{fig:effectofcell}. Another portion of the glass coupon was painted black on the backside, reflectance measurements over this part of the coupon are the ``ground truth" ARC measurement. When the probe is centered on a grid line, the reflectance spectrum includes metal reflectance and thus overestimates the true reflectance. However, when the probe is centered between grid lines, the reflectance spectrum is equal to the spectrum measured over the back-blackened section. Aligning the probe between grid lines can be accomplished with a steady hand by minimizing reflectance in real-time or by taking multiple measurements and determining alignment in post processing. Thus, a further advantage of the 1.0~mm probe is that by centering the probe between gridlines, reflectance from grid lines can be avoided.

The accuracy of the measurement in the UV ($< 400$~nm) can be compromised by stray light in the spectrometer~\cite{Zong2006}. Stray light artifacts are expected at wavelengths $< 400~\text{nm}$ because the tungsten halogen lamp (3100~K color temperature) has low output in this region. At these low wavelengths the photon flux falls off exponentially, yet a detection floor in the 200-300 nm region is observed due to stray light, see Fig.~\ref{fig:raw}. When the signal intensity becomes similar to the stray light intensity, either the reflectance measurement becomes invalid or some stray-light compensation is necessary~\cite{Zong2006}. For our specific setup (see Appendix~\ref{app:protocol}), we identify 400~nm as the lower limit for validity of the measurement.
 
One important limitation of the reflectance measurement is its behavior for soiled samples. Soiling that is not removed by gentle cleaning can both absorb and reflect light. Light reflected from soiling is directly measured by the reflectance probe, and the impact on light transmission to the cell can be accounted for. On the other hand, light absorbed by soiling is not measured in the reflectance spectrum; i.e. higher soil absorption reduces reflectance and/or transmittance. Fortunately, in many cases, the small 1.0~mm spot allows the user to choose a minimally soiled location.

For general ARC measurements, these experiments and calculations show that reducing the aperture diameter to 1.0~mm leads to an accurate measurement. This aperture size is also usually small enough to avoid metallization on the solar cell. Measurements with an aperture significantly smaller than 1.0 mm can suffer from inadequate light intensity and other optical difficulties.

\section{Equivalence of Probe Geometries}

\begin{figure}
\centering
\includegraphics{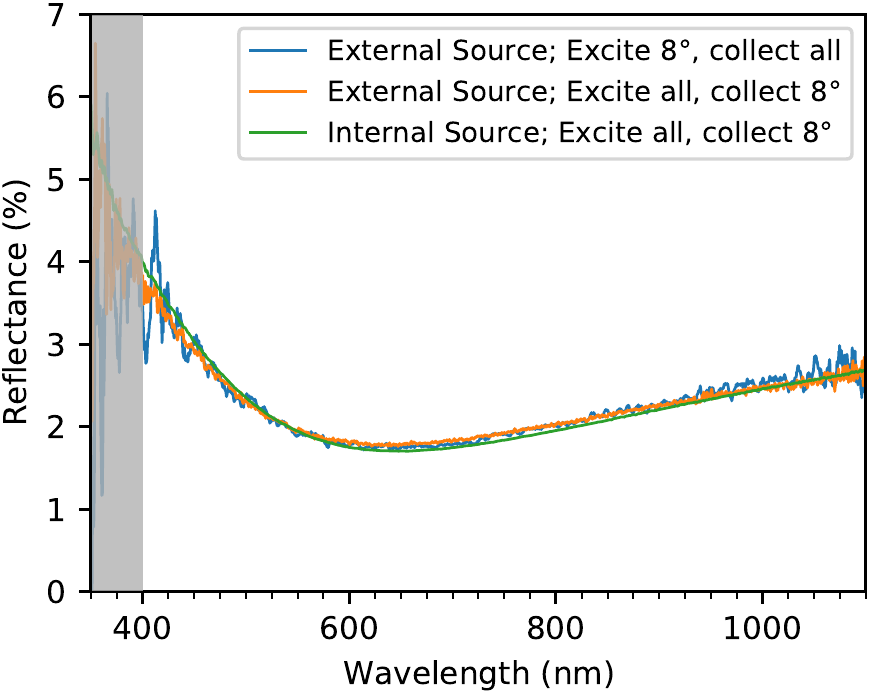}
\caption{Demonstration that reflectance spectra measured in either optical geometry give the same result, albeit with different signal to noise ratios due to the different light intensities. ``External source; excite all" geometry has a higher signal-to-noise ratio than ``external source; excite 8$^\circ$'' because a larger fiber is used, 0.64~mm diameter (Thorlabs BFL200HS02) compared to 1.3~mm diameter (Thorlabs BF13HSMA02) respectively. When using a probe with an internal light source (Ocean Insight ISP-REF), the light intensity is even greater and good signal to noise is observed from 400-1100 nm. Total exposure time for each spectrum was 0.6 seconds, although the higher-intensity spectra were acquired with multiple averages of shorter exposures. The ``external source" spectra used a boxcar averaging of 5 points, while ``internal source" did not use any averaging.
}
\label{fig:reciprocity} 
\end{figure}

Two related optical geometries can be used to measure the reflectance spectrum. Either light is injected into port A and collected from port B (excite 8$^\circ$, collect all angles) or light is injected into port B and collected from port A (excite all angles, collect 8$^\circ$), see Fig.~\ref{fig:diagram}. Conveniently, the Helmholtz reciprocity theorem implies that both measurements result in the same reflectance spectrum \cite{Clarke1985}. An example of this equivalence is demonstrated on a flat glass coupon with a porous silica ARC (unaged sample D from \cite{Miller2020}); chosen for its high spatial homogeneity. Both measurement geometries result in the same reflectance spectrum, see Fig.~\ref{fig:reciprocity}. Measurements on several other samples (not shown) also confirm the result. 

While either geometry results in the same reflectance spectrum, is is easier to achieve a high light intensity when exciting all angles because a light source can be placed inside the sphere (Fig.~\ref{fig:reciprocity} internal source).
%The ``excite all angles, collect 8$^\circ$" geometry results in a less noisy reflectance spectrum because greater light intensity is achievable.
%When collecting from port B, a 7-core round-to-linear fiber bundle (0.64~mm effective diameter, Thorlabs BFL200HS02) is used to couple light into the 100~$\mu$m spectrometer slit. When injecting into port B, a 19-core round-to-round fiber bundle is used instead (1.3~mm effective diameter, Thorlabs BF13HSMA02) since no spatial constraint is present on the light source side. 
Therefore, the ``excite all angles, collect 8$^\circ$" geometry is superior because the higher light intensity lowers photon noise and the impact of ambient light. In the following, we use a probe with an internal light source (Ocean Insight ISP-REF) which gave the best performance of the three optical configurations tested (Fig.~\ref{fig:reciprocity}).

\section{Theory of Single-Layer ARCs}\label{sec:theory}

While the measurement described in Sec.~\ref{sec:methods} is applicable to a general ARC, in this section we theoretically describe reflectance from a single-layer ARC with a uniform effective index of refraction. We model the spectral reflectance using the matrix form of the Fresnel equations following Ref.~\cite{TMM}. We consider a stack of three materials: a semi-infinite layer of air (material 0), a thin film of a given thickness (material 1) and a semi-infinite layer of glass (material 2). For a uniform thin film, the reflectance for unpolarized light is 
\begin{equation}\label{eq:R}
R = \frac{1}{2} \left| \frac{ e^{i \delta} r_{s,01} + e^{-i \delta} r_{s,12}}{ e^{-i \delta} + e^{i \delta} r_{s,01}  r_{s,12}} \right|^2 +  \frac{1}{2} \left| \frac{ e^{i \delta} r_{p,01} + e^{-i \delta} r_{p,12}}{ e^{-i \delta} + e^{i \delta} r_{p,01}  r_{p,12}} \right|^2
\end{equation}
where \({\delta = 2 \pi n_1 a \cos \theta_1 / \lambda}\), \(a\) is the thickness of the thin film, \(\theta_j\) is the angle of incidence of light in material \(j\) and \(r_{s/p,ij}\) is the reflectance amplitude for \(s/p\) polarized light from material \(i\) to material \(j\). The reflectance amplitudes are:
\begin{align}
	r_{s,ij} & = \frac{n_i \cos \theta_i  - n_j \cos \theta_j }{n_i \cos \theta_i  + n_j \cos \theta_j}\\
	r_{p,ij} & = \frac{n_j \cos \theta_i  - n_i \cos \theta_j }{n_j \cos \theta_i  + n_i \cos \theta_j },
\end{align} 
where \(n_j\) is the index of refraction of material \(j\). These equations give a convenient analytic form for predicting the spectral reflectance of an ARC. Since maximum destructive interference in Eq.~\ref{eq:R} occurs when \({\delta=\pi/2}\); the thickness of the ARC is typically tuned to a quarter wavelength.

The calculations require knowledge of the index of refraction of the various layers: air, porous silica ARC and glass. For a nano-porous material, volume-averaging effective-medium theory provides an effective index of refraction given the constituent parts:
\begin{equation}
n_\text{eff} = \sqrt{ (1 - P) n_c^2 + P n_d^2}
\end{equation}
where \(P\) is the porosity, \(n_c\) and \(n_d\) are the indices of refraction of the continuous phase and dispersed phases respectively~\cite{Braun2006}. The index of fused silica is used for the continuous phase~\cite{Malitson1965}, and air \({n_d = 1.00029}\) is used for the diffuse phase. For the module glass, we use the refractive index of soda-lime glass~\cite{Rubin1985}.

\begin{figure*}
\centering
\includegraphics{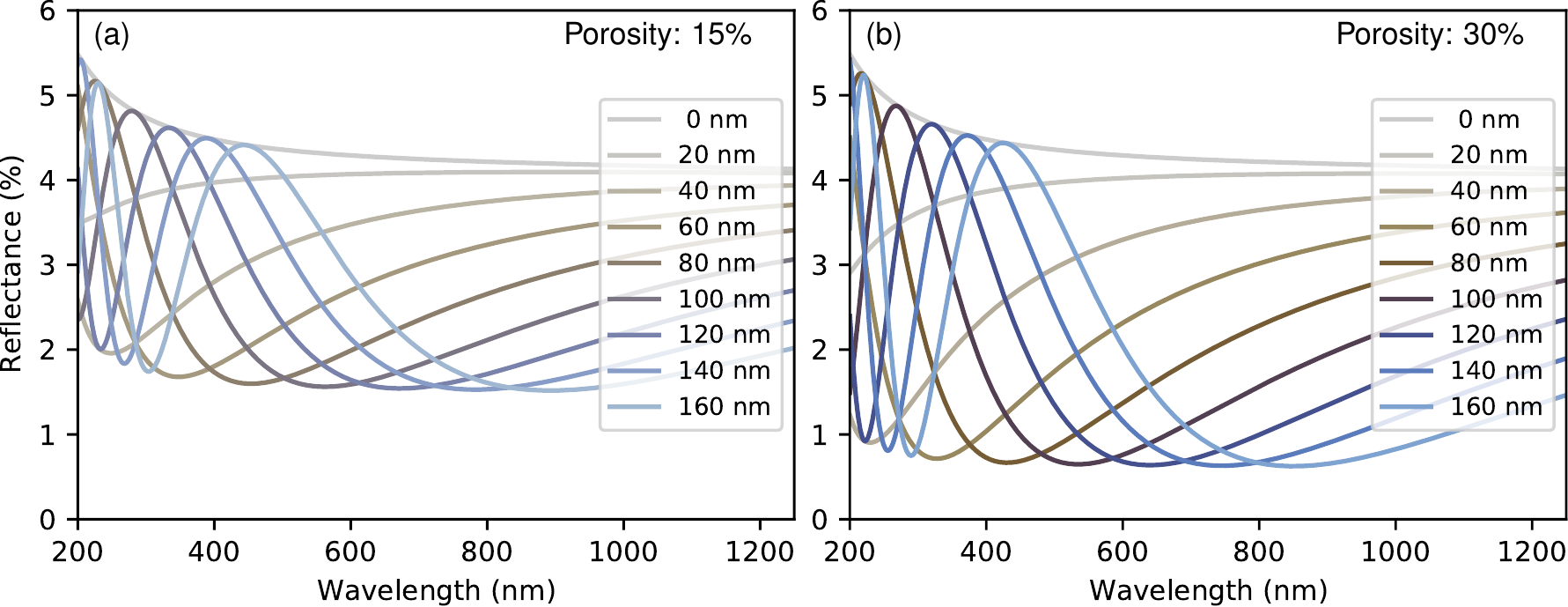}
\caption{Simulated spectral reflectance curves for a porous silica ARC on soda-lime glass in air at 8$^\circ$ angle-of-incidence. The thickness of the porous silica coating is varied in each plot. A destructive interference dip is observed that depends on the film thickness. As the porosity increases from 15\% (a) to 30\% (b), reflectance decreases further.}
\label{fig:thickness}
\end{figure*}

Using these refractive indices, the reflectance spectrum from a single-layer ARC can be calculated as a function of film thickness and porosity. The coating has a strong destructive interference dip where film thickness controls the dip center and porosity controls the dip depth, see Figure~\ref{fig:thickness}. Some manufacturers may use a graded-index ARC or multiple layers to lower reflectance further~\cite{Raut2011}, which could be treated with similar methods.

\section{ARC Performance Metrics}

A useful figure of merit for the coating performance is the solar-weighted photon reflectance (SWPR), the reflectance weighted by the AM1.5 photon flux~\cite{Miller2013}: 
\begin{equation}
\text{SWPR} = \frac{ \int_{\lambda_\text{min}}^{\lambda_\text{max}} R(\lambda) E_\text{p}(\lambda) d\lambda}{ \int_{\lambda_\text{min}}^{\lambda_\text{max}} E_\text{p}(\lambda) d\lambda},
\end{equation}
where \(E_\text{p}(\lambda) =  E (\lambda)\cdot \lambda/(h c)\), \(E(\lambda)\) is the AM1.5 spectrum in units of W/m$^2$/nm~\cite{AM1p5}, \(\lambda\) is the wavelength, \(h\) is Planck's constant and \(c\) is the speed of light.
Weighting by the photon flux is appropriate for solar photovoltaic applications because only one electron-hole pair is typically created from each photon. Integration limits from 300 to 1250 nm have been proposed as a standard because they include all of today's single-junction devices~\cite{Miller2013}, however reduced limits of 400 to 1100 nm are more appropriate for a mono-Si device and are used in this paper~\cite{Ganjoo2019}.

\begin{figure}
\centering
\includegraphics{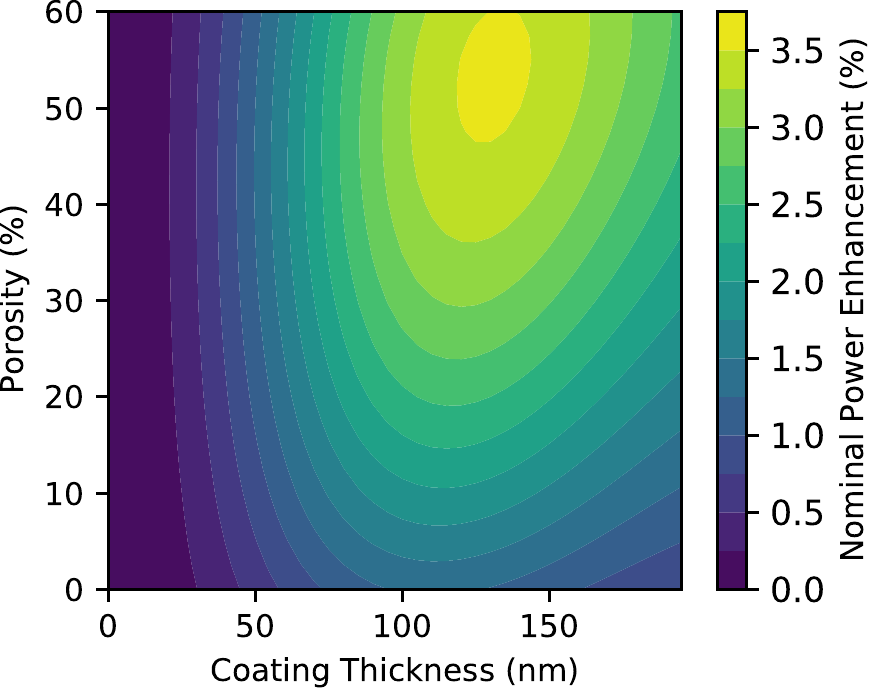}
\caption{Dependence of nominal power enhancement (NPE) with integration limits from 400 to 1100~nm on porosity and coating thickness at 8$^\circ$ angle-of-incidence. Higher-porosity coatings have better performance but can be less durable. Simulation calculates reflectance spectra for a coating of porous silica on soda-lime glass with 8$^\circ$ angle-of-incidence.}
\label{fig:NPE}
\end{figure}

Because the external quantum efficiency (EQE) for most commercial devices has a top-hat shape, we can define the nominal power enhancement (NPE) due to an ARC as:\begin{equation}
\text{NPE} = \text{SWPR}_\text{without-ARC} - \text{SWPR}_\text{with-ARC}
\end{equation}
where $\text{SWPR}_\text{without-ARC}$ is the SWPR for the air-glass interface without an ARC (4.2\% for soda-lime glass at $8^\circ$). We show the NPE for a layer of porous silica on soda-lime glass at 8$^\circ$ angle of incidence as a function of thickness and porosity in Fig.~\ref{fig:NPE}. As porosity increases, higher NPE is possible, however higher porosities can also compromise coating durability~\cite{Nielsen2015,Simurka2018}. For each porosity, an optimal coating thickness can be found that maximizes NPE, these values are shown in Table~\ref{tab:opt}. The actual energy yield enhancement due to an ARC for an operational system depends on device design and deployment conditions since single-layer ARCs improve light collection even more at higher angles of incidence~\cite{Pasmans2019} and some ARCs also have anti-soiling properties~\cite{Einhorn2019,Glaubitt2012}.

\begin{table}
\centering
\caption{Dependence of optimal coating thickness, nominal power enhancement (NPE) and solar-weighted photon reflectance (SWPR) on porosity for a single layer of porous silica on soda-lime glass at 8 degree angle-of-incidence. Integration limits for NPE and SWPR are 400 to 1100 nm. NPE and SWPR do not change by more than 0.01\% absolute when using 0 degree angle-of-incidence.}
\label{tab:opt}
\begin{tabular}{llll}
\toprule
Porosity & Max NPE & Min SWPR & Thickness (nm) \\
\midrule
      0\% &       1.30\% &        2.96\% &     110.9 \\
      5\% &       1.64\% &        2.61\% &     112.4 \\
     10\% &       1.97\% &        2.29\% &     114.2 \\
     15\% &       2.27\% &        1.98\% &     115.8 \\
     20\% &       2.55\% &        1.70\% &     117.5 \\
     25\% &       2.80\% &        1.45\% &     119.3 \\
     30\% &       3.03\% &        1.23\% &     121.2 \\
     35\% &       3.22\% &        1.04\% &     123.2 \\
     40\% &       3.37\% &        0.89\% &     125.3 \\
     45\% &       3.48\% &        0.78\% &     127.5 \\
     50\% &       3.54\% &        0.71\% &     129.8 \\
     55\% &       3.56\% &        0.70\% &     132.1 \\
     60\% &       3.51\% &        0.75\% &     134.7 \\
\bottomrule
\end{tabular}
\end{table}

\section{Model Fitting and Extrapolation}\label{sec:fit}

The model described in Sec.~\ref{sec:theory} can be used to estimate the coating thickness and porosity from the reflectance spectrum. Such interferometric models have accurately determined coating thickness and porosity in previous studies~\cite{Galy2020}. 
For a clean anti-reflective coating a simplified model of spectral reflectance uses Eq.~\ref{eq:Rtot} with
\begin{equation}\label{eq:Rmodel}
R_\text{air-glass} = \beta_\text{ARC} R_\text{ARC}(a, P) + (1- \beta_\text{ARC}) R_\text{ARC}(a=0,P)
\end{equation}
where \(\beta_\text{ARC}\) is the fractional area covered by the ARC (between 0 and 1) and $R_\text{ARC}(a, P) $ is the reflectance from the air-coating-glass interface calculated using the method in Sec.~\ref{sec:theory}. Although more complicated models could incorporate a range of thicknesses and porosities, Eq.~\ref{eq:Rmodel} is a simplified three-parameter model that describes the main degradation modes for ARCs: uniform thickness loss and deep scratches with complete coating loss. If an aperture smaller than 1.2~mm is used, then we typically approximate \({f_\text{cell}=f_\text{enc}=0}\) in Eq.~\ref{eq:Rtot}.

Because the ARC measurement is typically only valid within a certain wavelength limit, it is important to only use the valid portion of the reflectance spectrum for fitting the model. From measurements on several mono- and multi-Si modules and comparison with typical EQE curves \cite{Green2020}, conservative valid wavelength limits are 475 to 1000~nm for mono-Si and multi-Si for apertures less than 1.0~mm. This can be concluded from Fig.~\ref{fig:effectofaperture} and Fig.~\ref{fig:effectofcell} as the range where the measured reflectance with a 1.0~mm aperture agrees closely with the textured glass reflectance. While the exact valid wavelength limits depend on several factors including the cell reflectance, spectrometer stray-light management, ambient light conditions and sphere aperture, these limits should be generally applicable for silicon PV.

In order to calculate SWPR, it can be necessary to use the model to extrapolate the reflectance spectrum above and below the valid wavelength range. Another option is to limit the integration bounds to the valid wavelength range as this is also the range where the EQE is highest.

%\cite{Yoldas1980}

\section{Field Measurement on Commercial PV Module} 

\begin{figure}
\centering
\includegraphics{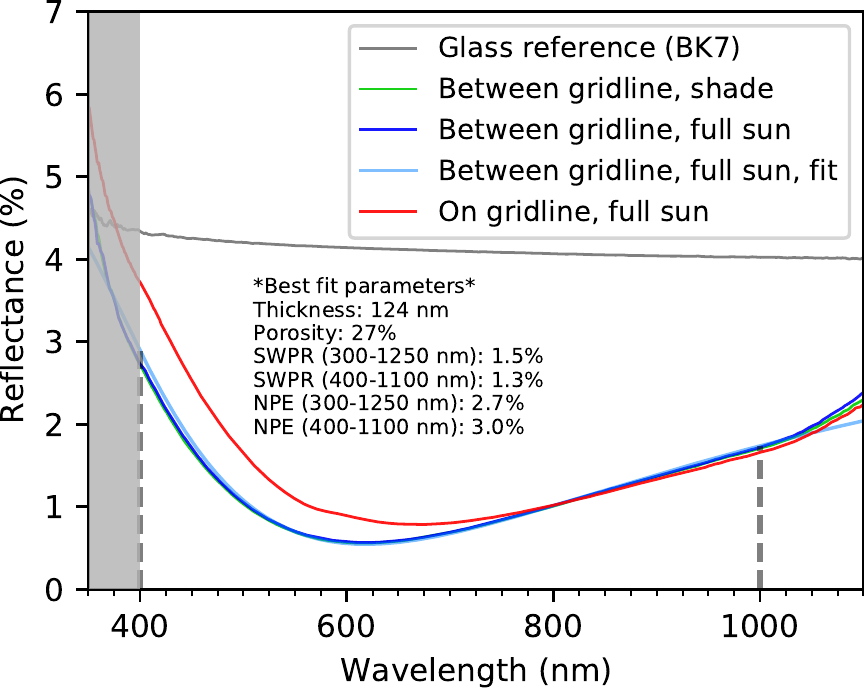}
\caption{Outdoor measurement of ARC on a Panasonic n-330 module. Spectral reflectance is lower when centering the probe between gridlines.  Uncertainty in measured SWPR and NPE can be estimated in several ways. The standard deviation of SWPR and NPE calculations found in nearby locations (closer than 2 mm) was 0.2\% absolute (data not shown). At the time of measurement, the module was in full sun. The small peak at 600~nm in the ``on gridline, full sun" measurement is due to red cemented debris on the module.}
\label{fig:field}
\end{figure}

The previously described methods can be used to characterize ARCs on active PV module installations. To demonstrate this, we measure the spectral reflectance of a Panasonic n330 module in outdoor conditions following the protocol in Appendix~\ref{app:protocol} using a 1.0~mm aperture, see Fig.~\ref{fig:field}. The measured reflectance when the sun is or is not blocked using a black cloth are very similar except at the band edge around 1100~nm. In this spectral region, ambient light makes a slight impact because it is coupled into the integrating sphere through multiple scattering events. Higher optical power or shading would further reduce the impact of ambient light. This comparison demonstrates that shading the probe from the sun has very little impact on the measured reflectance.

We used the method described in Sec.~\ref{sec:fit} to fit a model where the only adjustable parameters are the coating thickness and porosity (${\beta_\text{ARC}=1}$), finding excellent agreement with the data. The best-fit parameters give a coating porosity of 27\% and a thickness of 124~nm, close to the optimal thickness for this porosity (Tab.~\ref{tab:opt}). After 2.8 years of field aging in northern California with wet wipe cleaning once per year, these modules have retained an excellent ARC integrity with a nominal power enhancement of $3.0\%\pm0.2\%$. The uncertainty was found by repeating this measurement at multiple different nearby locations.

\section{Conclusion and Outlook}

This paper demonstrates techniques for accurately measuring glass coating reflectance on an installed PV module. Several main applications are anticipated.

First, there is a wide spread in ARC longevity suspected for commercial modules. This paper shows that non-destructive reflectance measurements can assess ARC degradation in fielded modules and provide feedback to PV system owners and module manufacturers. Knowledge of ARC degradation rates can also help predict future system performance. Because of the small interrogation area, the technique is also well suited to studying spatial variability in coating degradation.

Second, current PV product qualification programs do not test ARC longevity. With the techniques in this paper, a coating performance/durability test can be designed by measuring the glass reflectance before and after accelerated aging. Accelerated aging tests may target degradation caused by wet or dry abrasion (e.g. ISO 11998, DIN 53778-2, ASTM D2486 and DIN EN 1096-2) \cite{Ilse2019}, damp-heat, humidity-freeze or acid wash. 

This paper did not confirm whether the estimated thickness, porosity and fraction abraded agree with other measurements. Future work could use scanning electron microscopy and optical imaging to confirm the best-fit parameters \cite{Tamar2014}.

Last, we emphasize that although the reflectance measurement can be performed on clean or soiled PV modules, the estimation of a nominal power enhancement is only valid for clean modules. Future work could investigate other optical methods for estimating power loss due to soiling.

\appendices

\renewcommand\thefigure{\thesection.\arabic{figure}}    
\setcounter{figure}{0}    

\section{Details of Aperture Comparison}\label{app:aperture}

To compare the the reflectance spectrum from modules with different size apertures, we take measurements on a mono-Si minimodule  without an ARC on the glass. Apertures of various diameters were prepared using the method in App.~\ref{app:protocol}. Measurements were taken in the excite all angles, collect 8$^\circ$ geometry using a 30~mm diameter integrating sphere (Avasphere-30).The spectra were acquired following the protocol in App.~\ref{app:protocol}.  The focus of the collection fiber was not changed between measurements.   The textured glass sample was a piece of Swiftglass Solite glass with the backside spray painted black using four coats of a matte black spray paint (McMaster 7891T45).

%
%In order to acquire a reflectance spectrum, we acquire a background spectrum ($I_\text{dark}$) on a dark reference (Edmund optics 13-515), a reference spectrum ($I_\text{ref}$) using a flat BK7 glass coupon with a blackened back (Filmetrics REF-BK7) and a spectrum of the sample ($I_\text{sample}$) The reference reflectance ($R_\text{ref}$) is calculated using the Fresnel equations (XX ref). The sample reflectance spectrum is then:
%\begin{equation}
%R_\text{sample} = \frac{I_\text{sample} - I_\text{dark}}{ I_\text{ref} - I_\text{dark}}  R_\text{ref}
%\end{equation}
%

\section{Protocol for Measuring glass reflectance on assembled PV modules}\label{app:protocol}

In this section we provide a protocol for measuring air-glass reflectance from an assembled PV module. Part numbers used in this paper are listed; other equipment can also accomplish the same task. The commercial probes with the lowest photon noise and best ambient light rejection have an internal light source (Ocean Insight ISP-REF, AvaSphere-50-LS-HAL-12V). For indoor measurements, better performance in the UV could be achieved with a broadband (e.g. deuterium-halogen) light source and a separate integrating sphere (Avantes AvaSphere-30, Ocean Insight ISP-30-6-R). A further improvement to stray-light performance could be achieved with a color balancing filter (e.g. Thorlabs FGT05165 or Ocean Insight OF2-BG34R). 

First we describe how to prepare the measurement apparatus. This process takes several hours once the materials have been collected.

\begin{figure}
\centering
\includegraphics{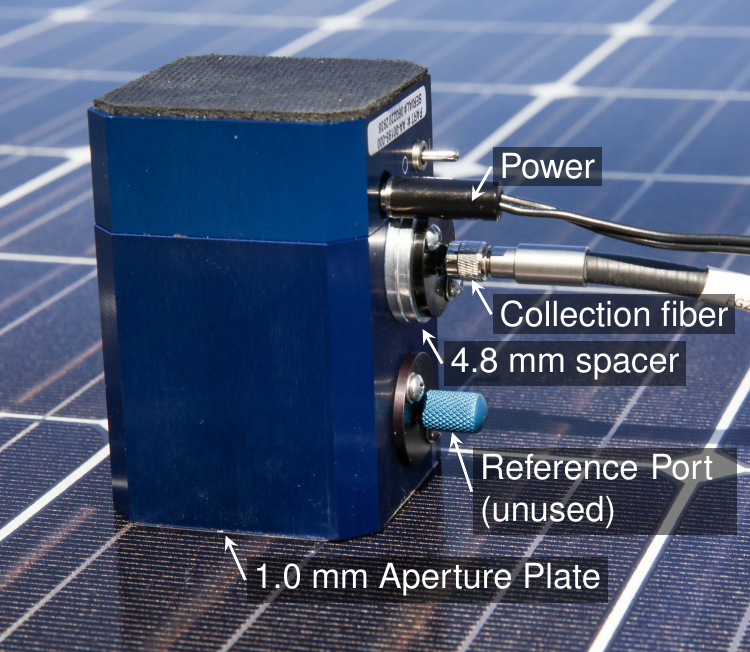}
\caption{Image of modified integrating sphere probe (Ocean Insight ISP-REF) on a PV module. The tungsten-halogen light source is inside the sphere. The modifications include fixing a stainless steel shim plate with a 1.0~mm hole between onto the probe and adding a spacer between the collection fiber and the body of the sphere. The spacer adjusts the distance between the collection fiber and an internal lens in order to focus onto the 1.0~mm interrogation area. }
\label{fig:probe}
\end{figure}

\begin{enumerate}
\item \textbf{Align lens.} Turn on a broadband light source (e.g. Ocean Insight HL-2000) and reduce the light intensity. Use a 200~$\mu$m single-core multimode fiber (Thorlabs FG200AEA, coated steel sleeve) to inject light into port A of an integrating sphere (Ocean Insight ISP-REF, port S). Place a piece of scotch tape over the sample aperture so the focused light spot can be observed in the plane of the aperture. Adjust the position of the port A fiber relative to the focusing lens to focus the light to a minimum spot size. Using Ocean Insight ISP-REF, focus by placing a spacer with 4.8~mm thickness between the probe body and the fiber port (2x Mcmaster 92510A677, with holes drilled to accommodate screws, replace screws with \#4-40 x 3/8 in), see Fig.~\ref{fig:probe}. 
\item \textbf{Prepare aperture.} Drill a 1.0 to 1.1 mm diameter hole in the center of a piece of 0.005~in. stainless steel shim stock with adhesive back (McMaster 2958A25, 2951N212). Trim the shim to fit the reflectance probe. Use matte black spray paint (McMaster 16625T411) to blacken the sample side of the aperture. Optionally apply PTFE tape (McMaster 6802K12) in an annulus around the aperture on the probe side to increase light intensity. 
\item \textbf{Attach and align aperture.} Place a piece of scotch tape on the sample side of the aperture, adjust aperture or fiber position in order to center the focused light through the aperture. Fix the aperture using the provided adhesive or tape and secure the fiber.
\item \textbf{Switch to excite all, collect 8$^\circ$ geometry.} Connect the 200~$\mu$m single-core fiber from port A to the spectrometer (Ocean HDX) using a large slit (100~$\mu$m). Inject broadband light into the sphere using an internal or external light source. If using Ocean Insight ISP-REF, set to include specular reflection.
\item \textbf{Prepare dark reference.} The dark reference can be a commercial light trap (Edmund optics 13-515) or a blackened box. To make the blackened box, open a $\sim$1~cm hole in a cardboard box larger than 6" square. Line the interior of the box with black aluminum foil (Thorlabs BKF12) or spray paint the interior of the box with matte black paint (e.g. McMaster 7891T45).
\item \textbf{Prepare light reference.} The suggested light reference is a piece of BK7 glass~\cite{schottBK7} with the back painted black (Filmetrics REF-BK7) or absorptive neutral density glass (Thorlabs NE80B or Ocean Insight STAN-SSL).
\end{enumerate}

The following protocol is used to clean the sample. 

\begin{enumerate}[resume]
\item \textbf{Clean sample.} Do not clean a module with cold water in full-sun conditions as this can heat shock a PV module. First, spray clean with deionized water (DI) and Liqui-Nox (Alconox Inc.) detergent solution and a spray bottle, without contacting the module. Blow dry the sample with air or nitrogen gas. If the glass is visibly clean then contact cleaning is not necessary. It is important to remove as much debris as possible using non-contact methods since any remaining debris could scratch the ARC during contact cleaning. If contact cleaning is needed, spray module with cleaning solution and use light pressure to wipe module with a cloth cleanroom wipe (Twillx 1622, Berkshire Corp., Kimwipe or similar) in one direction only. For subsequent strokes use a clean wipe each time. 
\item \textbf{Wait to dry.} The sample must be completely dry (including the internal pores) in order to get an accurate measurement, a standard wait time is 30 minutes. Some porous silica coatings absorb water into interior pores and take up to 30 minutes to dry completely in the lab. However, in direct sun on a warm day, several minutes of drying is often sufficient. The drying time can be identified for the current coating and conditions by repeatedly measuring reflectance (described below) until no change is observed. Blowing air can speed the drying process and improve cleaning.
\end{enumerate}

\begin{figure}
\centering
\includegraphics{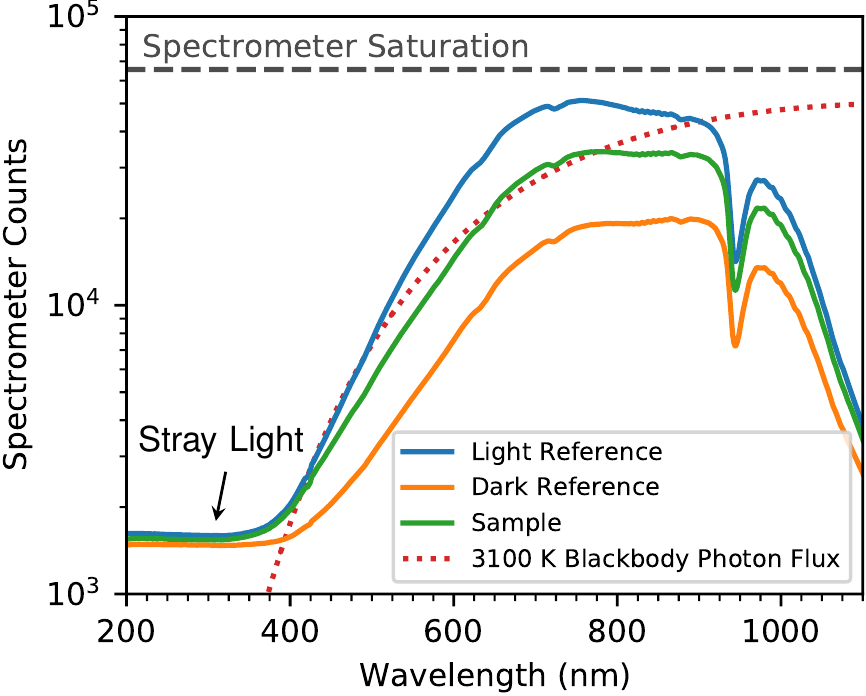}
\caption{Example raw spectrometer output for measurement in Fig.~\ref{fig:effectofcell} ``ARC over black paint."
Sample reflectance is calculated from the measured signal on the sample, light and dark reference using Eq.~\ref{eq:calc}. The shape of the measured spectrum depends on the light source (a tungsten-halogen bulb with a 3100~K color temperature), the optical system and the detector response. The dip at 950 nm is due to absorption in the collection fiber. The intensity fall-off above 1000~nm is due to the responsivity of the silicon detector. This measurement has a stray light floor around 450x lower than the spectral maximum. When the signal counts become similar to the stray light intensity ($< 400$~nm), either stray light compensation is needed or the reflectance measurement is no longer accurate. Each spectrum consists of 60 averages of 10~ms exposures.
}
\label{fig:raw}
\end{figure}

The reflectance spectrum is acquired using the following method.
\begin{enumerate}[resume]
\item \textbf{Stabilize light source.} Allow light source and spectrometer to warm up for the allotted time, typically 30 minutes. 
\item \textbf{Take baseline measurement (light reference).} Clean light reference with a glass cleaner and wipe. Place probe on light reference sample and adjust integration time and number of averages so the spectrometer is not saturated. We use a 6~ms exposure time and 50 averages with nonlinearity correction enabled. Store spectrum as $I_\text{ref}$. 
\item \textbf{Take 0\% baseline (dark reference).} Place the integrating sphere probe on the dark reference and store this spectrum as $I_\text{dark}$.  Considerable light intensity may still be measured due to internal reflections in the sphere. 
\item \textbf{Take measurement on sample.} If measuring an assembled module with front-side metallization, it is important to center the probe between grid lines. Carefully place the probe on the glass surface over the PV cell, making as minimal a change as possible to the bending of the optical fibers from the light/dark reference measurements. By hand, displace the probe slowly in the direction perpendicular to the grid lines while monitoring the spectrum, finding a location that minimizes the total reflectance spectrum. Store this spectrum as $I_\text{sample}$. Repeat in slightly different locations by displacing parallel to the grid lines in order to acquire more measurements. If real time feedback is not possible, take $>$20 measurements at slightly different locations and average the the spectra with total intensity lower than the 15th percentile. 
\item \textbf{Retake baseline measurement.} Repeat the light reference measurement and check whether the reflectance has shifted by more than 0.1\% absolute. If the light reference has shifted, the previous measurement is invalid and must be retaken under more stable conditions. Common reasons for instability are light-source power changes due to varying environmental conditions, unclean reference, changing ambient light intensity and fiber bending.
\item \textbf{Calculate absolute reflectance.} The sample reflectance can be calculated by
\begin{equation}\label{eq:calc}
R_\text{sample} = \frac{I_\text{sample} - I_\text{dark}}{ I_\text{ref} - I_\text{dark}}  R_\text{ref}.
\end{equation}
where $R_\text{ref}$ is the reflectance of the light reference sample calculated using the Fresnel equations~\cite{TMM}. Many spectrometer software applications can perform this calculation in real time. Better accuracy in dim spectral regions can be obtained by correcting for stray light artifacts~\cite{Zong2006}. 
\item \textbf{Additional considerations for outdoor measurements.} If working outside, protect the light source from changing sun and wind conditions in order to minimize the change in the probe spectrum. The AC output of a portable power source can be used to power the light source (Omnicharge Omni 20+). High solar insolation conditions can affect the reflectance spectrum if sunlight is coupled into the integrating sphere. At higher light levels, a black light-blocking cloth can be placed over the probe and module to block ambient light from entering the sphere. A quick comparison of the reflectance spectrum when the ambient light is blocked and unblocked can determine if the light block is necessary in the current conditions. 
\end{enumerate}

\section*{Acknowledgment}

The authors would like to thank Jenya Meydbray and Michael Owen-Bellini for providing mini-module specimens.

\ifCLASSOPTIONcaptionsoff
  \newpage
\fi



%\bibliographystyle{IEEEtran}
%\begin{thebibliography}{1}
%
%\end{thebibliography}

\bibliographystyle{IEEEtran}
\bibliography{toddkarin.bib}


%\vfill


\end{document}